\documentclass[preprint,3p,times,twocolumn,sort&compress]{elsarticle}

\usepackage{ecrc_nologo}

\runauth{M.\ Sjö et al.}

\jid{nppp}

\jnltitlelogo{Nuclear and Particle Physics Proceedings}

\usepackage{amssymb}

\usepackage[figuresright]{rotating}

\usepackage{amsmath}

\newcommand{\p}{\partial}

\newcommand{\T}{\mathsf{T}}

\DeclareMathOperator{\SU}{SU}

\DeclareMathOperator{\OO}{O}

\newcommand{\rr}{\mathrm{r}}
\newcommand{\lr}[1]{\ell_{#1}^\rr}

\newcommand{\Mphys}{M_\mathrm{phys}}
\newcommand{\Mpi}{M_\pi}
\newcommand{\Fphys}{F_\mathrm{phys}}
\newcommand{\Fpi}{F_\pi}

\newcommand{\df}{\mathrm{df}}

\newcommand{\Kdf}{\cK_{\df,3}}

\newcommand{\cD}{\mathcal{D}}

\newcommand{\cK}{\mathcal{K}}
\newcommand{\cL}{\mathcal{L}}
\newcommand{\cM}{\mathcal{M}}

\newcommand{\cO}{\mathcal{O}}

\newcommand{\lagr}{\cL}         
\newcommand{\ampl}{\cM}         

\renewcommand{\vec}[1]{\boldsymbol{#1}}

\newcommand{\LO}{\text{LO}}
\newcommand{\NLO}{\text{NLO}}
\newcommand{\Kiso}{{\cK_0}}
\newcommand{\Kisoone}{{\cK_1}}
\newcommand{\Kisotwo}{{\cK_2}}
\newcommand{\KA}{{\cK_\mathrm{A}}}
\newcommand{\KB}{{\cK_\mathrm{B}}}
\newcommand{\DeltaA}{{\Delta_\mathrm{A}}}
\newcommand{\DeltaB}{{\Delta_\mathrm{B}}}

\newcommand{\MF}[1]{\bigg(\frac{\Mpi}{\Fpi}\bigg)^{\!#1}}

\newcommand{\lrI}{\ell_1^\rr}
\newcommand{\lrII}{\ell_2^\rr}
\newcommand{\lrIII}{\ell_3^\rr}
\newcommand{\lrIV}{\ell_4^\rr}

\newcommand{\elliso}{{\ell_{(0)}^\rr}}
\newcommand{\ellisoone}{{\ell_{(1)}^\rr}}
\newcommand{\ellisotwo}{{\ell_{(2)}^\rr}}
\newcommand{\ellA}{{\ell_{(\mathrm{A})}^\rr}}
\newcommand{\ellB}{{\ell_{(\mathrm{B})}^\rr}}

\newcommand{\Diso}{{\cD_0}}
\newcommand{\Disoone}{{\cD_1}}
\newcommand{\Disotwo}{{\cD_2}}
\newcommand{\DA}{{\cD_\mathrm{A}}}
\newcommand{\DB}{{\cD_\mathrm{B}}}

\usepackage{hyperref}
\usepackage{cleveref}
\newcommand{\rcite}[1]{Ref.~\cite{#1}}
\newcommand{\rrcite}[1]{Refs.~\cite{#1}}

\hypersetup{%
    colorlinks,%
    breaklinks,%
    citecolor=blue,urlcolor=blue,linkcolor=blue,%
    hypertexnames=false,%
    pdftitle={Three-particle scattering: From the chiral Lagrangian to the lattice},%
    pdfauthor={Mattias Sjo}%
}

\usepackage{subcaption}
\usepackage{pgf,tikz}
\usetikzlibrary{decorations.pathreplacing,decorations.text,shapes,intersections}
\tikzset{
    prop/.style={thick,join=round}
}
\newcommand{\diagramscale}{.95}
\newcommand{\Bradius}{.4}
\newcommand{\Bdiam}{.8}
\newcommand{\Cradius}{.5}

\newcommand{\nonpole}{\text{non-pole}}

\newcommand{\looptools}{{\em LoopTools}}
\newcommand{\cpp}{C\texttt{++}}

\usepackage{pgfplots}
\pgfplotsset{compat=1.17}   
\usepgfplotslibrary{fillbetween}
\pgfmathsetlengthmacro\MajorTickLength{
      \pgfkeysvalueof{/pgfplots/major tick length} * 0.4
    }
\pgfmathsetlengthmacro\MinorTickLength{
      \pgfkeysvalueof{/pgfplots/minor tick length} * 0.3
    }
\tikzset{
    LOdash/.style={dash pattern=on 3pt off 2.5pt},
    fitdash/.style={dash pattern=on 1pt off 1.3pt},
    zeroline/.style={mark=none, black, ultra thin},
    verticalline/.style={thin, black, dash pattern=on 3pt off 3.5pt},
    KXline/.style={thick},
    DXline/.style={ultra thick},
    LOline/.style={KXline, black,   LOdash},
    NLOline/.style={KXline, fitgray},
    NLOupperline/.style={line width = 0pt, fitgray},
    NLOlowerline/.style={line width = 0pt, fitgray},
    Fitline/.style={KXline, fitorange, fitdash},
    Fitupperline/.style={line width = 0pt, fitorange},
    Fitlowerline/.style={line width = 0pt, fitorange},
    NLO2line/.style={NLOline},
    NLO2upperline/.style={NLOupperline},
    NLO2lowerline/.style={NLOlowerline},
    NLOAline/.style={NLOline, LOdash},
    NLOAupperline/.style={NLOupperline},
    NLOAlowerline/.style={NLOlowerline},
    numericline/.style={KXline, red},
    threshline/.style={KXline, blue, dash pattern=on 3pt off 2.5pt},
    D0line/.style={DXline, plotI,   solid},
    D1line/.style={DXline, plotII,  dash pattern=on 8pt off 2pt on 2pt off 2pt},
    D2line/.style={DXline, plotIII, dash pattern=on 6pt off 2pt on 2pt off 2pt on 2pt off 2pt},
    DAline/.style={DXline, plotIV,  dash pattern=on 4.4pt off 4.4pt},
    DBline/.style={DXline, plotV,   dash pattern=on 2pt off 2pt},
    allwaveline/.style={DXline, black},
    swaveline/.style={D1line},
    dwaveline/.style={DAline},
    gwaveline/.style={DBline},
}
\pgfkeys{/pgf/number format/1000 sep={}}
\pgfplotsset{
    general plot/.style={
        set layers=axis on top,
        width=1.0\textwidth,
        height=0.75\textwidth,
        minor tick num=4,
        every tick/.style={
                semithick,
            },
        major tick length=\MajorTickLength,
        minor tick length=\MinorTickLength,
        y tick label style={
                /pgf/number format/.cd,
                fixed,
             fixed zerofill,
                precision=1,
                /tikz/.cd
                },
            every tick label/.append style={font=\footnotesize},
            axis line style={black, semithick},
        xticklabel style={inner sep=1.5pt},
        yticklabel style={inner sep=1.5pt},
            },
    fit plot/.style={
        general plot,
        ylabel style = {rotate=-90}, ylabel shift=-1.1ex,
        legend style={font=\scriptsize}
    },
    legend image code/.code={
        \draw [#1] (0pt,0pt) -- (15pt,0pt);
    },
    filled legend/.style={legend image code/.code={
        \draw [#1] (0pt,0pt) -- (15pt,0pt);
        \path [draw=none, fill=#1, opacity=0.3] (0pt,-4pt) rectangle (15pt,4pt);
    }},
    double filled legend/.style={legend image code/.code={
        \draw [#1] (0pt,5pt) -- (15pt,5pt);
        \path [draw=none, fill=#1, opacity=0.3] (0pt,1pt) rectangle (15pt,9pt);
        \draw [#1, solid] (0pt,-5pt) -- (15pt,-5pt);
        \path [draw=none, fill=#1, opacity=0.3] (0pt,-1pt) rectangle (15pt,-9pt);
    }},
    legend with mark/.style={legend image code/.code={
        \draw [#1, thick, solid] (7.5pt,4pt) -- (7.5pt,-4pt);
        \draw [#1, thick, solid, fill=white] (7.5pt,0pt) circle[radius=1.5pt];
    }},
    filled legend with mark/.style={legend image code/.code={
        \draw [#1] (0pt,0pt) -- (15pt,0pt);
        \path [draw=none, fill=#1, opacity=0.3] (0pt,-4pt) rectangle (15pt,4pt);
        \draw [#1, thick, solid] (7.5pt,4pt) -- (7.5pt,-4pt);
        \draw [#1, thick, solid, fill=white] (7.5pt,0pt) circle[radius=1.5pt];
    }},
    wide legend/.style={legend image code/.code={
        \draw [#1] (0pt,0pt) -- (22pt,0pt);
    }},
    lattice data/.style={
        thick,
        scatter, scatter/use mapped color={draw=#1,fill=white},
        only marks, mark=*,
        mark options={scale=0.7},
        error bars/.cd, y dir = both, x dir = both, y explicit, y explicit, error bar style={color=#1, thick, mark size = 0pt},
        /pgfplots/.cd
    },
    text along plot/.style 2 args={
        mark=none, draw=none,
        decoration={text along path, text align=center, raise=1ex,
            text={#1},
            text align={left indent=#2} },
        postaction={decorate}}
}

\definecolor{fitorange}{HTML}{FFA500}
\definecolor{fitblue}{HTML}{0000FF}
\definecolor{fitgray}{HTML}{808080}
\definecolor{plotI}{HTML}{0077BB}
\definecolor{plotII}{HTML}{33BBEE}
\definecolor{plotIII}{HTML}{009988}
\definecolor{plotIV}{HTML}{EE7733}
\definecolor{plotV}{HTML}{CC3311}

\begin{document}

\begin{frontmatter}



\dochead{}

\title{Three-pion scattering: From the chiral Lagrangian to the lattice\tnoteref{fund}\tnoteref{preprint}}

\tnotetext[fund]{This document is the result of several research projects funded by a large number of US and EU grants, listed at the end of \rcite{Baeza:2023ljl}.}
\tnotetext[preprint]{Preprint: MIT-CTP/5598, LU TP 23-11}


\author[a]{Jorge Baeza-Ballesteros}
\author[b]{Johan Bijnens}
\author[b,c]{Tom\'a\v s Husek}
\author[d]{\\Fernando Romero-L\'opez}
\author[e]{Stephen R.\ Sharpe}
\author[b]{Mattias Sjö\corref{speaker}}

\address[a]{IFIC, CSIC-Universitat de València, 46980 Paterna, Spain}
\address[b]{Department of Physics, Lund University, Box 118, SE 22100 Lund, Sweden}
\address[c]{Institute of Particle and Nuclear Physics, Charles University, V Hole\v sovi\v ck\'ach 2, 180 00 Prague, Czech Republic}
\address[d]{CTP, Massachusetts Institute of Technology, Cambridge MA 02139, USA}
\address[e]{Physics Department, University of Washington, Seattle, WA 98195-1560, USA}

\cortext[speaker]{Speaker, corresponding author.}

\begin{abstract}
    In recent years, detailed studies of three-pion systems have become possible in lattice QCD.
    This has in turn led to interest in 3-to-3 scattering of pions in the chiral perturbation theory framework.
    In addition to being an interesting study of multi-meson dynamics in its own right, it provides a valuable handle on finite-volume effects and the pion mass dependence, thus complementing the lattice results.
    I present our derivation of the next-to-leading order amplitude for this process, as well as its conversion into the three-particle K-matrix, which enables direct comparison to the lattice.
    Our results significantly improve the agreement between theory and lattice, which was poor when only leading-order effects were taken into account.
\end{abstract}

\begin{keyword}
Chiral Lagrangian \sep Hadronic Spectroscopy \sep Structure and Interactions \sep Lattice QCD


\end{keyword}

\end{frontmatter}


\section{Introduction}

Lattice QCD and chiral perturbation theory (ChPT) are two widely used tools for low-energy QCD.
While the lattice provides a general first-principles numerical approach, ChPT allows perturbative calculation of many processes and can be used to control certain systematics of the lattice, such as finite-volume effects and pion mass dependence.
Thus, recent advances in lattice QCD (see references in \rcite{Baeza:2023ljl}, particularly \rrcite{Rusetsky:2019gyk,Hansen:2019nir}) have caused interest in the study of 3-to-3 scattering using ChPT, a process that formerly had only been studied at leading order \cite{Osborn:1969ku,Susskind:1970gf}; the higher-order tree-level counterterms were also recently computed \cite{Low:2019ynd,Bijnens:2019eze}.

These proceedings are partly based on \rcite{Bijnens:2021hpq}, where we performed the first one-loop determination of the 3-to-3 pion scattering amplitude, and its follow-up \rcite{Bijnens:2022zsq}, where we generalized that result to cases including more than 2 quark flavors; however, results for which $m_s\neq m_{u,d}$ are still unavailable.
It is also based on \rcite{Baeza:2023ljl}, in which this amplitude is converted into the 3-particle K-matrix, a scheme-dependent object that allows the finite-volume energy spectrum to be determined.
The K-matrix can also be calculated on the lattice (as of this writing, \rcite{Blanton:2021llb} has the most precise values), allowing for comparison to ChPT.
Previous results using leading-order ChPT \cite{Blanton:2019vdk} disagree strongly with the lattice results.

\section{The $3\pi\to3\pi$ amplitude in ChPT}
ChPT \cite{Gasser:1983yg,Gasser:1984gg} is the effective field theory that arises from the breaking of chiral symmetry down to its diagonal subgroup, giving the breaking pattern \mbox{$\SU(n)\times\SU(n)/\SU(n)$} for $n$ quark flavors, as used in \rcite{Bijnens:2022zsq}.
For $n=2$, an equivalent formulation is $\OO(4)/\OO(3)$, which is used in \rcite{Bijnens:2021hpq}.

In the $\OO(4)/\OO(3)$ formulation, the Lagrangian is
\begin{equation}\label{eq:lagr}
    \begin{aligned}
        \lagr &= \tfrac{F^2}{2} \p_\mu \Phi^\T\p^\mu\Phi + F^2\chi^\T\Phi\\
            &+ \ell_1 (\p_\mu\Phi^\T\p^\mu\Phi)(\p_\nu\Phi^T\p^\nu\Phi)\\
            &+ \ell_2 (\p_\mu\Phi^\T\p_\nu\Phi)(\p^\mu\Phi^T\p^\nu\Phi)\\
            &+ \ell_3 (\chi^\T\Phi)^2
            + \ell_4 \p_\mu\chi^\T\p^\mu\Phi + \ldots\,,
    \end{aligned}
\end{equation}
where the first line is LO, and orders above NLO have been omitted.
Here, $\ell_i$ are (bare) coupling constants, \mbox{$\chi^\T = (M^2, \vec 0)$} with $M$ the (bare) pion mass, and $\Phi$ is a 4-component vector that can be parametrized in terms of the pion fields in multiple ways; the one used in \rcite{Baeza:2023ljl} was introduced by Weinberg \cite{Weinberg:1968de} and is
\begin{equation}
    \Phi = \frac{1}{1 + \vec\phi^\T\vec\phi/4F^2}\left(1 - \vec\phi^\T\vec\phi/4F^2, \tfrac{1}{F}\vec\phi^\T\right)^\T\,,
\end{equation}
where $\vec\phi^\T=(\phi_1,\phi_2,\pi^0)$, $\pi^\pm = \frac{1}{\sqrt2}(\phi_1\mp i\phi_2)$,
and $F$ is the (bare) pion decay constant.
See \rcite{Bijnens:2021hpq} for a more extensive description of the parametrizations, and \rcite{Bijnens:2022zsq} for a novel derivation of the most general parametrization for $n\geq 2$.

The couplings $\ell_i$ are renormalized to $\lr{i}$ using
\begin{equation}\label{eq:lr}
    \ell_i = -\kappa\frac{\gamma_i}{2}\left[\tfrac1\epsilon - \log\tfrac{\mu^2}{4\pi} + 1\right] + \lr{i}
\end{equation}
in $4-2\epsilon$ dimensions at $\mu=770~\text{MeV}\approx M_\rho$, with
\begin{equation}
    \gamma_1 = \tfrac13\,,\quad
    \gamma_2 = \tfrac23\,,\quad
    \gamma_3 = \tfrac13\,,\quad
    \gamma_4 = 2
\end{equation}
and $\kappa\equiv1/(16\pi^2)$.
$M,F$ are renormalized to $M_\pi,F_\pi$.

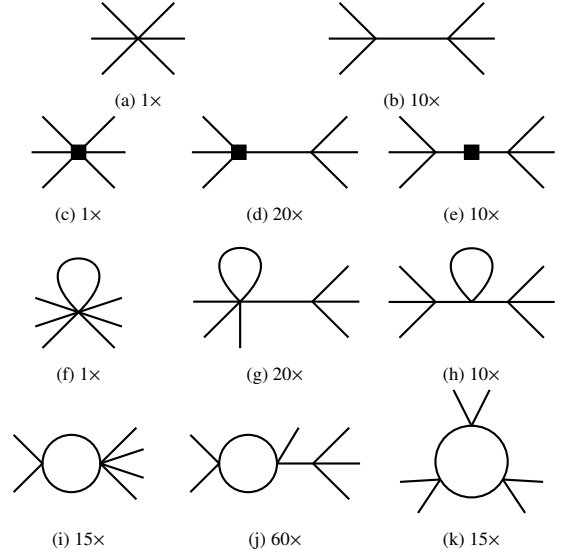
\begin{figure}[t]
    \centering
    \begin{subfigure}[t]{0.45\columnwidth}
        \centering
        \begin{tikzpicture}[scale=\diagramscale]
            \draw [prop] (-.5,.5)   -- (0,0) -- (-.65,0);
            \draw [prop] (-.5,-.5)  -- (0,0) -- (.5,-.5);
            \draw [prop] (.65,0)     -- (0,0) -- (.5,.5);
        \end{tikzpicture}
        \caption{$1\times$}
        \label{fig:6piLO:point}
    \end{subfigure}
    \begin{subfigure}[t]{0.45\columnwidth}
        \centering
        \begin{tikzpicture}[scale=\diagramscale]
            \draw [prop] (-.5,.5)   -- (0,0) -- (-.65,0);
            \draw [prop] (-.5,-.5)  -- (0,0) -- (1,0) -- (1.5,-.5);
            \draw [prop] (1.65,0)    -- (1,0) -- (1.5,.5);
        \end{tikzpicture}
        \caption{$10\times$}
        \label{fig:6piLO:prop}
    \end{subfigure}

    \begin{subfigure}[t]{0.32\columnwidth}
        \centering
        \begin{tikzpicture}[scale=\diagramscale]
            \draw [prop] (.65,0)    -- (0,0) -- (-.65,0);
            \draw [prop] (-.5,.5)   -- (0,0) -- (.5,.5);
            \draw [prop] (-.5,-.5)  -- (0,0) -- (.5,-.5);
            \draw [fill=black] (-.1,-.1) rectangle (.1,.1);
        \end{tikzpicture}
        \caption{$1\times$}
        \label{fig:6piNLO:point}
    \end{subfigure}
    \begin{subfigure}[t]{0.32\columnwidth}
        \centering
        \begin{tikzpicture}[scale=\diagramscale]
            \draw [prop] (-.5,.5)   -- (0,0) -- (-.65,0);
            \draw [prop] (-.5,-.5)  -- (0,0) -- (1,0) -- (1.5,-.5);
            \draw [prop] (1.65,0)    -- (1,0) -- (1.5,.5);
            \draw [fill=black] (-.1,-.1) rectangle (.1,.1);
        \end{tikzpicture}
        \caption{$20\times$}
        \label{fig:6piNLO:prop}
    \end{subfigure}
    \begin{subfigure}[t]{0.32\columnwidth}
        \centering
        \begin{tikzpicture}[scale=\diagramscale]
            \draw [prop] (-.5,.5)   -- (0,0) -- (-.65,0);
            \draw [prop] (-.5,-.5)  -- (0,0) -- (1,0) -- (1.5,-.5);
            \draw [prop] (1.65,0)    -- (1,0) -- (1.5,.5);
            \draw [fill=black] (.5,0) +(-.1,-.1) rectangle +(.1,.1);
        \end{tikzpicture}
        \caption{$10\times$}
        \label{fig:6piNLO:2prop}
    \end{subfigure}

    \begin{subfigure}[t]{0.32\columnwidth}
        \centering
        \begin{tikzpicture}[scale=\diagramscale]
            \draw [white] (.65,0)    -- (0,0) -- (-.65,0);
            \draw [prop] (-.6,-.2)   -- (0,0) -- (.6,-.2);
            \draw [prop] (-.6,.2)   -- (0,0) -- (.6,.2);
            \draw [prop] (-.5,-.5)  -- (0,0) -- (.5,-.5);
            \draw [prop] (0,0) .. controls (-1,1) and (1,1) .. (0,0);
        \end{tikzpicture}
        \caption{$1\times$}
        \label{fig:6piNLO:A-point}
    \end{subfigure}
    \begin{subfigure}[t]{0.32\columnwidth}
        \centering
        \hskip-3mm
        \begin{tikzpicture}[scale=\diagramscale]
            \draw [prop] (-.65, 0) -- (0,0) -- (0,-.65);
            \draw [prop]  (0,0) .. controls +(-1,1) and +(1,1) .. +(0,0);
            \draw [prop] (-.5,-.5)   -- (0,0) -- (1,0) -- (1.5,-.5);
            \draw [prop] (1.65,0)    -- (1,0) -- (1.5,.5);
        \end{tikzpicture}
        \caption{$20\times$}
        \label{fig:6piNLO:A-prop}
    \end{subfigure}
    \begin{subfigure}[t]{0.32\columnwidth}
        \centering
        \begin{tikzpicture}[scale=\diagramscale]
            \draw [prop] (-.5,.5)   -- (0,0) -- (-.65,0);
            \draw [prop] (-.5,-.5)  -- (0,0) -- (1,0) -- (1.5,-.5);
            \draw [prop] (1.65,0)    -- (1,0) -- (1.5,.5);
            \draw [prop] (.5,0)  .. controls +(-1,1) and +(1,1) .. +(0,0);
            \draw [white] (0,-.5) -- (0,-.65);
        \end{tikzpicture}
        \caption{$10\times$}
        \label{fig:6piNLO:A-2prop}
    \end{subfigure}

    \begin{subfigure}[t]{0.32\columnwidth}
        \centering
        \begin{tikzpicture}[scale=\diagramscale]
            \draw [prop] (-.4,.4)   -- (0,0) -- (-.4,-.4);
            \draw [prop] (\Bradius,0) circle[radius=\Bradius];
            \begin{scope}[shift={(\Bdiam,0)}]
                \draw [prop] (.5,.5) -- (0,0) -- (.5,-.5);
                \draw [prop] (.6,.2) -- (0,0) -- (.6,-.2);
            \end{scope}
            \draw[white] (0,-.5) -- (0,-.7);
        \end{tikzpicture}
        \caption{$15\times$}
        \label{fig:6piNLO:B-point}
    \end{subfigure}
    \begin{subfigure}[t]{0.32\columnwidth}
        \centering
        \begin{tikzpicture}[scale=\diagramscale]
            \draw [prop] (-.4,.4)   -- (0,0) -- (-.4,-.4);
            \draw [prop] (\Bradius,0) circle[radius=\Bradius];
            \begin{scope}[shift={(\Bdiam,0)}]
                \draw [prop] (.3,.5) -- (0,0) -- (1.15,0);
                \draw [prop] (1,.5) -- (.5,0) -- (1,-.5);
            \end{scope}
            \draw[white] (0,-.5) -- (0,-.7);
        \end{tikzpicture}
        \caption{$60\times$}
        \label{fig:6piNLO:B-prop}
    \end{subfigure}
    \begin{subfigure}[t]{0.32\columnwidth}
        \centering
        \begin{tikzpicture}[scale=\diagramscale]
            \draw [prop] (0,0) circle[radius=\Cradius];
            \foreach \x in {0,120,240} {
                \begin{scope}[rotate=\x]
                    \draw [prop] (0,\Cradius) -- +(.25,.5);
                    \draw [prop] (0,\Cradius) -- +(-.25,.5);
                \end{scope}
            }
        \end{tikzpicture}
        \caption{$15\times$}
        \label{fig:6piNLO:C}
    \end{subfigure}
    \caption{
        LO (top row) and NLO Feynman topologies for the six-point amplitude.
        Black squares indicate NLO vertices; remaining vertices are LO.
        The number of distinct diagrams obtainable through crossing from each topology is indicated below it.}
    \label{fig:6pi}
\end{figure}

\Cref{fig:6pi} shows the Feynman diagrams relevant for the six-point (i.e., 3-to-3) amplitude, $\ampl_3$.
The amplitude partly factorizes over the propagator pole:
\begin{equation}\label{eq:factor}
    \ampl_3 = -\sum \frac{\ampl_2^{(L)} \ampl_2^{(R)}}{b^2 - M^2_\pi} + \ampl_3^{(\nonpole)}\,,
\end{equation}
where $\ampl_{2}^{(L,R)}$ are 4-point amplitudes with one leg off-shell, corresponding to the left and right vertices of a diagram like \cref{fig:6piLO:prop}; $\pm b$ is the momentum of those off-shell legs, and of the propagator that joins them; the sum is over the ways of distributing the 6 on-shell external legs between $\ampl_2^{(L,R)}$; and $\ampl_3^{(\nonpole)}$ is whatever remains of the full amplitude.
Such a pole is present in \cref{fig:6piLO:prop,fig:6piNLO:prop,fig:6piNLO:2prop,fig:6piNLO:A-prop,fig:6piNLO:A-2prop,fig:6piNLO:B-prop}, but the correspondence is neither exact nor unique:
The contribution of each diagram depends on the parametrization, and the structure of \cref{eq:factor} depends on the convention used for off-shell amplitudes.

Most of the non-factorizing part of the calculation follows as a simple extension of the 4-point amplitude derived in \rcite{Bijnens:2011fm}, since \cref{fig:6piLO:point,fig:6piNLO:point,fig:6piNLO:B-point} are essentially 4-point diagrams with extra legs on some vertices.
Only the triangle loop diagram, \cref{fig:6piNLO:C}, presents a problem, since conventional Passarino--Veltman reduction results in extremely long expressions.
We found it necessary to devise a redundant basis of more symmetric triangle loop functions, labelled $C$, $C_{11}$, $C_{21}$ and $C_3$, which are listed in appendix~A of \rcite{Bijnens:2021hpq}.
In terms of these, $\ampl_3^{(\nonpole)}$ is listed in appendix~B of \rcite{Bijnens:2021hpq} or, in a different and more general form, appendix~D of \rcite{Bijnens:2022zsq}.

\section{The $3\pi$ K-matrix}
Following the formalism of \rcite{Hansen:2014eka}, the K-matrix, $\Kdf$, determines the finite-volume energy spectrum $\{E_n\}$ of three pions in a box of size $L$ through the quantization condition
\begin{equation}\label{eq:quant}
    \det\Big[ F_3^{-1}(E, \vec P, L) + \Kdf(E^*) \Big] = 0 \quad\text{at $E = E_n$}\,.
\end{equation}
Here, $(E,\vec P)\equiv P$ is the total 4-momentum, $E^*\equiv \sqrt{P^2}$ is the corresponding center-of-momentum energy.
$F_3$ is described in \rcite{Hansen:2014eka}.
\Cref{eq:quant} generalizes Lüscher's
2-particle quantization condition \cite{Luscher:1986pf,Luscher:1990ux}.

In this formalism, $\Kdf$ is Lorentz-invariant and constructed entirely from on-shell quantities; it does, however, contain a scheme-dependent cutoff function.
The subscript ``df'' indicates that it is divergence-free: Unlike $\ampl_3$, it has neither poles nor cuts.
In general, the relation between $\Kdf$ and $\ampl_3$ is a complicated integral equation, but at NLO in the chiral expansion, it reduces to an algebraic subtraction.
Importantly, $\Kdf$ is purely real, so the subtraction of cuts can be wholly circumvented by dropping the imaginary parts.
We only give a qualitative description of the subtraction here; the derivation and detailed form can be found in \rcite{Baeza:2023ljl}.

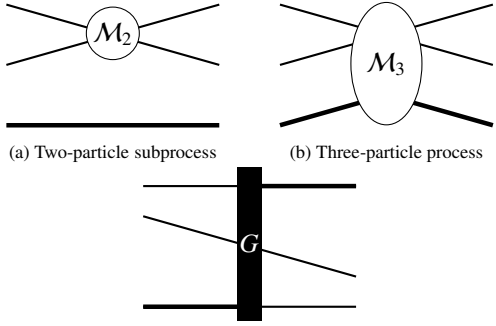
\begin{figure}[tbp]
    \centering
    \begin{subfigure}[t]{0.45\columnwidth}
        \centering
        \begin{tikzpicture}[scale=2]
            \draw [thick] (-.7,+.2) -- (0,0) -- (+.7,+.2);
            \draw [thick] (-.7,-.2) -- (0,0) -- (+.7,-.2);
            \draw [line width=.6mm] (-.7,-.6) -- (+.7,-.6);

            \node [shape=circle,draw=black,fill=white,inner sep=1pt,anchor=mid] at (0,0) {$\ampl_2$};


        \end{tikzpicture}
        \caption{Two-particle subprocess}
        \label{fig:schem:2pi}
    \end{subfigure}
    \begin{subfigure}[t]{0.45\columnwidth}
        \centering
        \begin{tikzpicture}[scale=2]
            \draw [thick] (-.7,+.2) -- (0,0) -- (+.7,+.2);
            \draw [thick] (-.7,-.2) -- (0,0) -- (+.7,-.2);
            \draw [line width=.6mm] (-.7,-.6) -- (0,-.4) -- (+.7,-.6);

            \node [shape=ellipse,draw=black,fill=white,inner xsep=2pt,inner ysep=12pt,anchor=mid] at (0,-.2) {$\ampl_3$};


        \end{tikzpicture}
        \caption{Three-particle process}
        \label{fig:schem:3pi}
    \end{subfigure}
    \\
    \begin{subfigure}[t]{\columnwidth}
        \centering
        \begin{tikzpicture}[scale=2]
            \draw [thick] (-.7,+.4) -- (+.7,+.4);
            \draw [thick] (-.7,-.4) -- (+.7,-.4);
            \draw [thick] (-.7,+.2) -- (+.7,-.2);
            \draw [line width=.6mm] (-.7,-.4) -- (0,-.4);
            \draw [line width=.6mm] (+.7,+.4) -- (0,+.4);

            \node [shape=rectangle,fill=black,inner xsep=1pt,inner ysep=.9cm,anchor=mid] at (0,0) {\color{white}$G$};

        \end{tikzpicture}
        \caption{The spectator-swapping function $G^\infty$. The diagonal line corresonds to the propagator being cancelled.}
        \label{fig:schem:G}
    \end{subfigure}
    \caption{Schematic representation of the components of $\Kdf$.
        Bold lines represent spectators, and thinner lines represent interacting pairs.}
    \label{fig:schem}
\end{figure}

Despite being a 3-particle quantity, a large part of $\Kdf$ is determined by 2-particle processes.
In light of this, we subdivide the three initial (or final) particles into a scattering pair and a non-scattering spectator; see \cref{fig:schem:2pi}.
The pair kinematics are decomposed into partial waves, allowing the two-particle amplitude to be expressed as
\begin{equation}
    \ampl_2(\vec p)_{\ell'm';\ell m}\,,
\end{equation}
where $\ell m$ ($\ell'm'$) are the inital-(final-)state partial wave indices, and $\vec p$ is the spectator momentum; everything is on-shell, so it is sufficient to use 3-momenta.
Supplemented with the total momentum $P$ (left implicit), this completely describes the kinematics.

We adopt the same description of 3-particle processes, and retain the notion of a spectator even though it is involved in the scattering; see \cref{fig:schem:3pi}.
(The particle exchange symmetries of the initial and final states can, in fact, be rephrased in terms of the freedom to choose a spectator.)
With $\vec p$ ($\vec p'$) as the initial-(final-)state spectator momentum, the 3-particle amplitude is
\begin{equation}
    \ampl_3(\vec p',\vec p)_{\ell'm';\ell m}\,.
\end{equation}
$\Kdf$ is expressed in the same form.

The last ingredient in $\Kdf$ is $G^\infty$, which is diagrammatically described in \cref{fig:schem:G}.
It serves to swap spectators between subsequent processes, and provides the mechanism for cancelling divergences.
Its general form is
\begin{equation}
    G^\infty(\vec p',\vec p) \sim \frac{H(x_{p'}) H(x_p)}{b_{p'p}^2 - \Mpi^2 + i\epsilon}\,,
\end{equation}
where $b_{p'p} \equiv P-p'-p$, and $H(x)$ is a cutoff function described below.
(We have omitted barrier factors that ensure smoothness and correct partial-wave behavior.)
Note how $G^\infty$ is similar in form to a propagator carrying momentum $b_{p'p}$; at the pole of the propagaor, they match exactly and cancel.
However, $G^\infty$ only connects on-shell quantities, and differs significantly from a propagator away from the pole.

The cutoff function $H(x)$, with $x_p \equiv (P-p)^2/(4\Mpi^2)$, may be any smooth function such that $H(x)=0$ for \mbox{$x<0$} and $H(x)=1$ for $x>1$; necessarily, such a function will be non-analytic.
It ensures that subtraction terms containing $G^\infty$ vanish far from the divergences they cancel, so that the subtraction does not introduce new UV-behavior on top of the already UV-finite amplitude.
The choice of $H$ constitutes the scheme-dependence of $\Kdf$;
the standard choice is
\begin{equation}\label{eq:H}
    H(x) = \exp\Big[-\tfrac1x \exp\Big(-\tfrac{1}{1-x}\Big)\Big]\,,\quad 0<x<1\,.
\end{equation}
Other choices and their effects on $\Kdf$ are studied in appendix~A of \rcite{Baeza:2023ljl}.

\section{The calculation of $\Kdf^\NLO$}
We work in the maximum-isospin channel (e.g., the process $3\pi^+\to 3\pi^+$), since it is structurally the simplest and the only one for which lattice data are currently available.
We also work in the threshold expansion, where the kinematics are expressed in terms of
\begin{equation}\label{eq:Delta}
    \begin{gathered}
        \Delta \equiv \frac{P^2 - 9\Mpi^2}{9\Mpi^2}\,,\qquad
        \Delta_i^{(\prime)} \equiv \frac{\big(P - p_i^{(\prime)}\big)^2 - 4\Mpi^2}{9\Mpi^2}\,,\\
        \tilde t_{ij} \equiv \frac{(p'_i - p_j)^2}{9\Mpi^2}\,,
    \end{gathered}
\end{equation}
which all vanish as $\cO(\Delta)$ at the 3-pion threshold.
The maximum-isospin channel expands as \cite{Blanton:2019igq}
\begin{equation}
    \Mpi^2\Kdf = \Kiso + \Kisoone \Delta + \Kisotwo \Delta^2 + \KA \DeltaA + \KB \DeltaB + \cO(\Delta^3)\,,
    \label{eq:threxp}
\end{equation}
where
\begin{equation}
    \DeltaA \equiv \sum_i\left(\Delta_i^2+\Delta_i^{\prime 2}\right)-\Delta^2\,,
    \qquad
    \DeltaB \equiv \sum_{i,j} \tilde t_{ij}^{\,2} -\Delta^2\,.
\end{equation}
It thus remains to compute the five coefficients $\cK_X$ with \mbox{$X=0,1,2,\mathrm A,\mathrm B$},
of which only the last two are sensitive to the angular distribution of the particles.

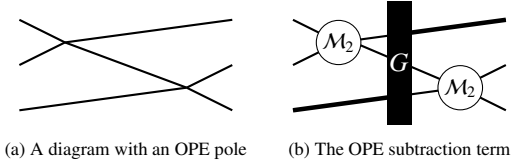
\begin{figure}[tbp]
    \centering
    \begin{subfigure}[c]{0.45\columnwidth}
        \centering
        \begin{tikzpicture}[yscale=1.5, xscale=2]
            \node [draw=none,shape=rectangle,inner xsep=1pt,inner ysep=.7cm,anchor=mid] at (0,0) {\color{white}$G$};

            \draw [thick] (-.4,+.2) -- (+.4,-.2);

            \draw [thick] (-.7,+.4) -- (-.4,+.2);
            \draw [thick] (-.7,+.0) -- (-.4,+.2);
            \draw [thick] (-.4,+.2) -- (+.7,+.4);

            \draw [thick] (+.7,-.4) -- (+.4,-.2);
            \draw [thick] (+.7,-.0) -- (+.4,-.2);
            \draw [thick] (+.4,-.2) -- (-.7,-.4);
        \end{tikzpicture}
        \caption{A diagram with an OPE pole}
        \label{fig:OPE:M}
    \end{subfigure}
    \begin{subfigure}[c]{0.45\columnwidth}
        \centering
        \begin{tikzpicture}[yscale=1.5, xscale=2]
            \draw [thick] (-.3,+.2) -- (+.4,-.2);

            \draw [thick] (-.7,+.4) -- (-.4,+.2);
            \draw [thick] (-.7,+.0) -- (-.4,+.2);
            \draw [thick, name path=upper] (-.4,+.2) -- (+.7,+.4);

            \draw [thick] (+.7,-.4) -- (+.4,-.2);
            \draw [thick] (+.7,-.0) -- (+.4,-.2);
            \draw [thick, name path=lower] (+.4,-.2) -- (-.7,-.4);

            \path [name path=vert] (0,+.5) -- (0,-.5);
            \draw [line width=.6mm, name intersections={of=upper and vert}] (intersection-1) -- (+.7,+.4);
            \draw [line width=.6mm, name intersections={of=lower and vert}] (intersection-1) -- (-.7,-.4);

            \foreach \x/\y in {-.4/+.2,+.4/-.2} {
                \node [shape=circle,draw=black,fill=white,inner sep=1pt,anchor=mid] at (\x,\y) {\footnotesize$\ampl_2$};
            }
            \node [shape=rectangle,fill=black,inner xsep=1pt,inner ysep=.7cm,anchor=mid] at (0,0) {\color{white}$G$};
        \end{tikzpicture}
        \caption{The OPE subtraction term}
        \label{fig:OPE:D}
    \end{subfigure}
    \caption{The OPE (one-particle exchange) pole and its subtraction, drawn schematically as in \cref{fig:schem}.
        The $s$-channel version of (a) is not present at maximum isospin.}
    \label{fig:OPE}
\end{figure}

At LO, the only divergence is the OPE (one-particle exchange) pole, exemplified by \cref{fig:OPE:M}, which is removed in $\Kdf$ by subtracting the term schematically given in \cref{fig:OPE:D}.
That is to say,
\begin{equation}
    \Kdf^\LO(\vec p',\vec p) = \ampl_3^\LO - \ampl^\LO_2(\vec p')G^\infty(\vec p',\vec p)\ampl^\LO_2(\vec p)\,,
    \label{eq:OPE}
\end{equation}
leaving the indices implicit.
$\Kdf^\LO$ is not scheme-dependent, since the cutoff functions are identically $1$ in the OPE subtraction.
The LO calculation was done already in \rcite{Blanton:2019vdk} and is reproduced in \rcite{Baeza:2023ljl}.

At NLO, the OPE subtraction must be augmented by promoting either 2-particle amplitude in \cref{eq:OPE} to NLO.
This introduces many more terms and partial waves, and requires great care in performing the calculations.
The subtraction is matched to the factorizing part of $\ampl_3$, given in \cref{eq:factor}, using the off-shell convention of \rcite{Bijnens:2021hpq}.
See sec.~4.4 of \rcite{Baeza:2023ljl} for more details.

\begin{figure}[tbp]
    \centering
    \begin{subfigure}[c]{0.35\columnwidth}
        \centering
        \begin{tikzpicture}[yscale=1.5, xscale=1]
            \node [draw=none,shape=rectangle,inner xsep=1pt,inner ysep=.7cm,anchor=mid] at (0,0) {\color{white}$G$};

            \draw [thick] (-.8,+.2) -- (0,-.2) -- (+.8,+.2) -- cycle;

            \draw [thick] (-1.1,+.4) -- (-.8,+.2);
            \draw [thick] (-1.1,+.0) -- (-.8,+.2);
            \draw [thick] (0,-.2) -- (-1.1,-.4);
            \draw [thick] (+1.1,+.4) -- (+.8,+.2);
            \draw [thick] (+1.1,+.0) -- (+.8,+.2);
            \draw [thick] (0,-.2) -- (+1.1,-.4);
        \end{tikzpicture}
        \caption{A bull's head diagram}
        \label{fig:BH:M}
    \end{subfigure}
    \begin{subfigure}[c]{0.55\columnwidth}
        \centering
        \begin{tikzpicture}[yscale=1.5, xscale=2]
            \draw [thick] (-.8,+.2) -- (0,-.2) -- (+.8,+.2);
            \draw [thick, name path=loop] (-.8,+.2) -- (+.8,+.2) node[midway, above] {\footnotesize$\vec r$};

            \draw [thick] (-1.1,+.4) -- (-.8,+.2);
            \draw [thick] (-1.1,+.0) -- (-.8,+.2);
            \draw [thick, name path=final] (0,-.2) -- (-1.1,-.4);
            \draw [thick] (+1.1,+.4) -- (+.8,+.2);
            \draw [thick] (+1.1,+.0) -- (+.8,+.2);
            \draw [thick, name path=initial] (0,-.2) -- (+1.1,-.4);

            \foreach \x/\y in {-.8/+.2,+.8/+.2,0/-.2} {
                \node [shape=circle,draw=black,fill=white,inner sep=1pt,anchor=mid] at (\x,\y) {\footnotesize$\ampl_2$};
            }

            \path [name path=vert] (-.4,+.5) rectangle (+.4,-.5);
            \draw [line width=.6mm, name intersections={of=final and vert}] (intersection-1) -- (-1.1,-.4);
            \draw [line width=.6mm, name intersections={of=initial and vert}] (intersection-1) -- (+1.1,-.4);
            \draw [line width=.6mm, name intersections={of=loop and vert}] (intersection-1) -- (intersection-2);

            \node [shape=rectangle,fill=black,inner xsep=1pt,inner ysep=.7cm,anchor=mid] at (-.4,0) {\color{white}$G$};
            \node [shape=rectangle,fill=black,inner xsep=1pt,inner ysep=.7cm,anchor=mid] at (+.4,0) {\color{white}$G$};
        \end{tikzpicture}
        \caption{The bull's head subtraction}
        \label{fig:BH:D}
    \end{subfigure}
    \caption{The ``bull's head'' triangle loop and its subtraction, drawn schematically as in \cref{fig:schem}.
        The version of (a) with its ``horns'' crossed, so that each vertex connects to one initial- and one final-state particle, is finite and lacks a corresponding subtraction term.}
    \label{fig:BH}
\end{figure}
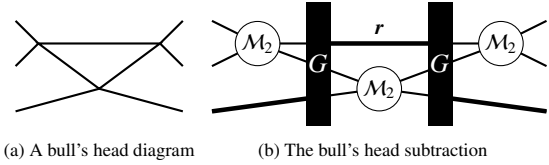

NLO also introduces a cut due to the triangle diagram, \cref{fig:6piNLO:C} or \ref{fig:BH:M}, which is subtracted by what we call the ``bull's head'' subtraction, \cref{fig:BH:D}, equal to
\begin{equation}
    -\int_r \ampl_2^\LO(\vec p') G^\infty(\vec p',\vec r) \ampl_2^\LO(\vec r) G^\infty(\vec r,\vec p)\ampl_2^\LO(\vec p)\,,
\end{equation}
where the integral is over all on-shell momenta $\vec r$.
Since we are only considering the finite real part, this can be treated separately from the triangle loop.

The cutoff functions involving the on-shell loop momentum $\vec r$ are not identically $1$, and this non-analyticity makes the integral rather challenging.
The most fruitful approach is to threshold-expand the integrand before integration, after which the angular part of the integral can be easily performed.
This involves expanding $H(x_r)$ in the vicinity of its essential singularities at $x_r=0$ and $1$, but this can be shown to be valid, since the derivatives of $H(x_r)$ vanish exponentially in this region.

These operations leave integrals of the type
\begin{equation}
     H_{m,n} \equiv \frac{1}{\pi^2}\int_0^{1/\sqrt{3}}\mathrm d z~\frac{\sqrt{1+z^2}}{z^{m}} \frac{\mathrm d^n}{\mathrm d x_r^n}\big[H^2(x_r)\big]\,,
    \label{eq:defHn}
\end{equation}
where $x_r = 1-3z^2$.
These possess endpoint singularities, which are intractable to the usual principal-value approach but can be regularized with Hadamard finite-part integration; for sufficiently smooth integrands, it works also when these singularities are essential \cite{Costin:2014hfp}.

The regularized integrals can be remarkably closely approximated by setting $H=1$ in \cref{eq:defHn}, giving easy-to-evaluate integrals plus small remainders to compute numerically.
These remainders encapsulate all the scheme-dependence in $\Kdf$.
See sec.~4.3 of \rcite{Baeza:2023ljl} for more details, including several complementary methods.

The last part of the calculation is to threshold-expand $\ampl_3^{\text{non-pole}}$.
Its real part can be extracted using Cauchy principal values, as described in sec.~4.2 of \rcite{Baeza:2023ljl}.

$\Kdf$ is the sum of the threshold-expanded amplitude, the OPE subtraction, and, at NLO, the bull's head subtraction.
We have performed several individual derivations and cross-checks of each part, using Wolfram Mathematica or FORM~\cite{Vermaseren:2000nd} for analytic calculations, and Mathematica or \cpp\ with CHIRON~\cite{Bijnens:2014gsa}, \looptools~\cite{Hahn:1998yk} and GSL for the numerics.

\section{Results and comparison to the lattice}

The LO contributions to $\Kdf$ are
\begin{equation}
    \Kiso       \supset 18\, \MF4\,,\qquad
    \Kisoone    \supset 27\, \MF4\,.
    \label{eq:results-LO}
\end{equation}
The quadratic-order terms in the threshold expansion vanish at LO.
The NLO contributions are [suppressing an overall factor of $(\Mpi/\Fpi)^6$]
\begin{equation}
    \begin{aligned}
        \Kiso
            &\supset \biggl[- 3\kappa(35 + 12\log3) - \Diso + 111L + \elliso\biggr]\,,
            \\
        \Kisoone
            &\supset \biggl[ - \frac{\kappa}{20}(1999 + 1920\log3) - \Disoone + 384L + \ellisoone \biggr]\,,
            \\
        \Kisotwo
            &\supset \biggl[\frac{207\kappa}{1400}(2923 - 420\log3) -\Disotwo + 360L +  \ellisotwo\biggr]\,,
            \\
        \KA
            &\supset \biggl[\frac{9\kappa}{560}(21809 - 1050\log3) - \DA - 9L + \ellA\biggr]\,,
            \\
        \KB
            &\supset \biggl[\frac{27\kappa}{1400}(6698 - 245\log3) - \DB + 54L + \ellB\biggr]\,,
    \end{aligned}
    \label{eq:results-NLO}%
\end{equation}
where $L\equiv\kappa\log(\Mpi^2/\mu^2)$ with $\mu$ and $\kappa$ described below \cref{eq:lr}, and $\ell^\rr_{(X)}$ contain the couplings, namely
\begin{equation}
    \begin{gathered}
        \elliso     = -288\lrI-432\lrII-36\lrIII+72\lrIV\,, \\
        \ellisoone  = -612\lrI-1170\lrII+108\lrIV\,, \quad
        \ellisotwo  = -432\lrI-864\lrII\,, \\
        \ellA       = 27\lrI+\tfrac{27}{2}\lrII\,, \quad
        \ellB       = -162\lrI-81\lrII\,,
    \end{gathered}
    \label{eq:LECthrexp}
\end{equation}
where we use the phenomenological values \cite{Colangelo:2001df, FLAG:2021npn}
\begin{equation}
    \begin{alignedat}{2}
        \bar\ell_1 &= -0.4(6)\,,\qquad& \bar\ell_2 &= 4.3(1)\,,\\
        \bar\ell_3 &= 3.07(64)\,,\qquad& \bar\ell_4 &= 4.02(45)
    \end{alignedat}
    \label{eq:LECref}
\end{equation}
throughout.
Lastly, $\cD_X$ are the small cutoff-dependent remainders from the bull's head subtraction, whose values using the standard cutoff choice, \cref{eq:H}, are
\begin{equation}
    \begin{gathered}
        \Diso       \approx -0.0563\,,\qquad
        \Disoone    \approx 0.130\,,\qquad
        \Disotwo    \approx 0.432\,,\\
        \DA         \approx 9.07\cdot10^{-4}\,,\qquad
        \DB         \approx 1.62\cdot10^{-4}\,.
    \end{gathered}
    \label{eq:Dthrexp}
\end{equation}
Their relative size compared to $\cK_X$, evaluated at the physical pion mass, range from $\sim{}10\%$ for $X=2$ to less than $0.1\%$ for $X=\mathrm A$.
$\cD_X$ stay consistently small for a wide selection of cutoff functions, as long as they are not too sharp.

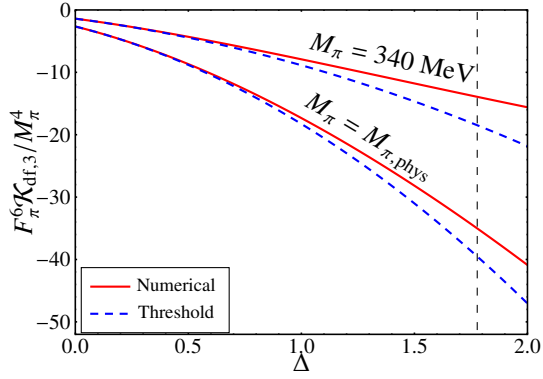
\begin{figure}[tb]
    \centering
    \begin{tikzpicture}[scale=.4]
        \begin{axis}[
            general plot,
            xmin=-0., xmax=2., xtick distance=0.5, restrict x to domain=-1:3,
            ymin=-52, ymax=0, ytick distance=10, restrict y to domain=-60:19,
            xlabel={$\Delta$},
            ylabel={$\Fpi^6\Kdf/\Mpi^4$},
            y tick label style={
                /pgf/number format/.cd,
                fixed,
                fixed zerofill,
                precision=0,
                /tikz/.cd},
            x tick label style={
                /pgf/number format/.cd,
                fixed,
                fixed zerofill,
                precision=1,
                /tikz/.cd},
            ylabel shift=2ex,
            legend style={font=\scriptsize, at={(0.03,0.35)},anchor=west}
            ]

            \addplot[numericline, mark=none]
                table[x index=0, y index=1, col sep=tab]
                {FitData/data340.txt};
            \addplot[threshline, mark=none]
                table[x index=0, y index=2, col sep=tab]
                {FitData/data340.txt};
            \addplot[numericline, mark=none]
                table[x index=0, y index=1, col sep=tab]
                {FitData/dataPP.txt};
            \addplot[threshline, mark=none]
                table[x index=0, y index=2, col sep=tab]
                {FitData/dataPP.txt};

            \addplot[text along plot={$M{_\pi}$ = 340 MeV}{1.7cm}]
                table[x index=0, y index=1, col sep=tab]
                {FitData/data340rough.txt};

            \addplot[text along plot={$M{_\pi}$ = $M{_\pi}{_,}{_\text{phys}}$}{0.7cm}]
                table[x index=0, y index=1, col sep=tab]
                {FitData/dataPPrough.txt};

            \addplot[verticalline] coordinates {(1.77778,19) (1.77778,-55)};

            \legend{Numerical, Threshold}

        \end{axis}
    \end{tikzpicture}
    \caption{
        Comparison between numerical results and the threshold expansion  for $\Kdf$, evaluated using  a fixed kinematic configuration described in \rcite{Baeza:2023ljl}, as a function of $\Delta$ defined in \cref{eq:Delta}.
        The comparison is presented for $\Mpi=\Mphys$  and $\Mpi=340$\;MeV, the latter corresponding to the heaviest pion mass used in \rcite{Blanton:2021llb}.
        The dashed vertical line indicates the inelastic threshold, which occurs at $E^*=5 \Mpi$.}
    \label{fig:threxp}
\end{figure}

As shown in \cref{fig:threxp}, the threshold expansion is in good agreement with the exact result up to the 5-pion threshold, were the $\Kdf$ formalism breaks down.

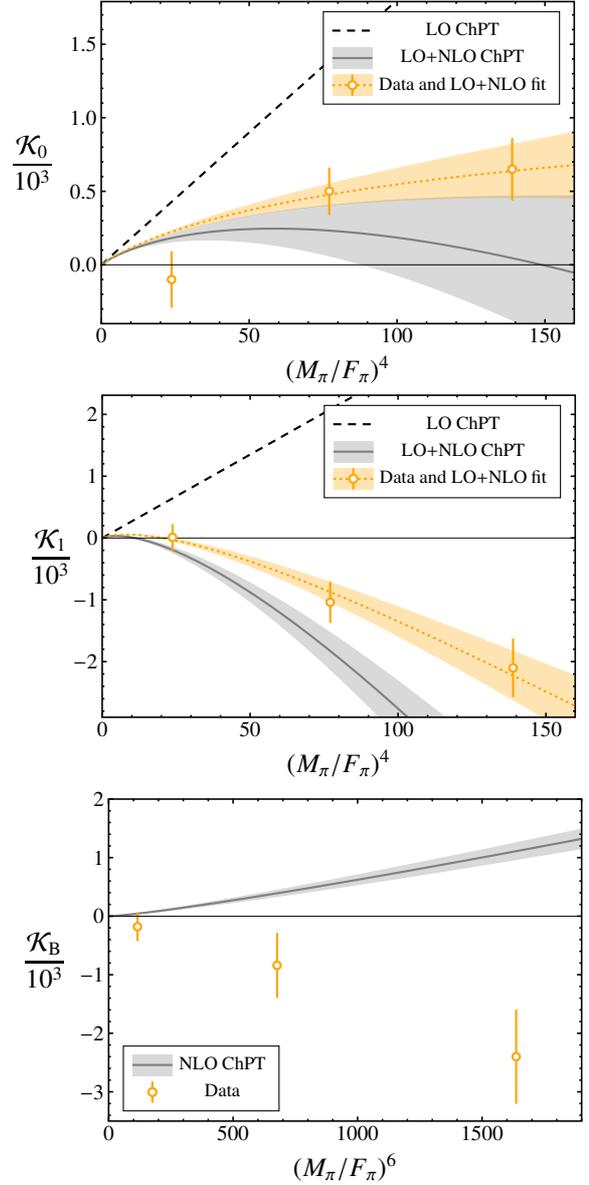
\begin{figure}[t!]
    \centering
    \begin{subfigure}[c]{\columnwidth}
        \begin{tikzpicture}
            \begin{axis}[
         		fit plot,
                xlabel={$(\Mpi/\Fpi)^4$},
                ylabel={$\displaystyle\frac{\Kiso}{10^3}$}, ylabel shift=0.2ex,
                ymin=-.4, ymax=1.79, ytick distance=.5, restrict y to domain=-1:2,
                xmin=-0.01, xmax=159.9, xtick distance=50, restrict x to domain=-1:180,
                legend pos = north east
                ]

                \addplot[LOline, mark=none]
                    table[x index=0, y index=1, col sep=tab]
                    {FitData/K0data.txt};

                \addplot[NLOline, mark=none, filled legend]
                    table[x index=0, y index=2, col sep=tab]
                    {FitData/K0data.txt};
                \addplot[NLOupperline, mark=none, name path global=NLOU,draw=none]
                    table[x index=0, y index=3, col sep=tab]
                    {FitData/K0data.txt};
                \addplot[NLOlowerline, mark=none, name path global=NLOL,draw=none]
                    table[x index=0, y index=4, col sep=tab]
                    {FitData/K0data.txt};

                \addplot[Fitline, mark=none, filled legend with mark]
                    table[x index=0, y index=5, col sep=tab]
                    {FitData/K0data.txt};
                \addplot[Fitupperline, mark=none, name path global=FitU,draw=none]
                    table[x index=0, y index=6, col sep=tab]
                    {FitData/K0data.txt};
                \addplot[Fitlowerline, mark=none, name path global=FitL,draw=none]
                    table[x index=0, y index=7, col sep=tab]
                    {FitData/K0data.txt};

                \addplot [fitorange, opacity=0.3] fill between[of=FitU and FitL];
                \addplot [fitgray, opacity=0.3] fill between[of=NLOU and NLOL];

    			\addplot [lattice data=fitorange]
               	table [x index = 0, y index = 1, x error index = 2, y error index = 3]
               	{FitData/K0points.txt};

               	\addplot[zeroline] coordinates {(-1,0) (170,0)};

                \legend{LO ChPT,LO+NLO ChPT,,,{Data and LO+NLO fit}}

    		\end{axis}
        \end{tikzpicture}
    \end{subfigure}%
    \\
    \begin{subfigure}[c]{\columnwidth}
        \centering
        \begin{tikzpicture}
            \begin{axis}[
         		fit plot,
                xmin=-0.01, xmax=159.9, xtick distance=50, restrict x to domain=-1:180,
                ymin=-2.9, ymax=2.31, ytick distance=1, restrict y to domain=-4:4,
                xlabel={$(\Mpi/\Fpi)^4$},
                ylabel={$\displaystyle\frac{\Kisoone}{10^3}$},
                legend pos = north east,
                y tick label style={
                /pgf/number format/.cd,
                fixed,
                precision=0,
                /tikz/.cd
                },
                ]

                \addplot[LOline, mark=none]
                    table[x index=0, y index=1, col sep=tab]
                    {FitData/K1data.txt};

                \addplot[NLOline, mark=none, filled legend]
                    table[x index=0, y index=2, col sep=tab]
                    {FitData/K1data.txt};
                \addplot[NLOupperline, mark=none, name path global=NLOU,draw=none]
                    table[x index=0, y index=3, col sep=tab]
                    {FitData/K1data.txt};
                \addplot[NLOlowerline, mark=none, name path global=NLOL,draw=none]
                    table[x index=0, y index=4, col sep=tab]
                    {FitData/K1data.txt};

                \addplot[Fitline, mark=none, filled legend with mark]
                    table[x index=0, y index=5, col sep=tab]
                    {FitData/K1data.txt};
                \addplot[Fitupperline, mark=none, name path global=FitU,draw=none]
                    table[x index=0, y index=6, col sep=tab]
                    {FitData/K1data.txt};
                \addplot[Fitlowerline, mark=none, name path global=FitL,draw=none]
                    table[x index=0, y index=7, col sep=tab]
                    {FitData/K1data.txt};

                \addplot [fitorange, opacity=0.3] fill between[of=FitU and FitL];
                \addplot [fitgray, opacity=0.3] fill between[of=NLOU and NLOL];

                \addplot [scatter, only marks, mark=*, scatter/use mapped color={draw=fitorange,fill=orange}, mark options={scale=0.7}]
               	table [x index = 0, y index = 1]
               	{FitData/K1points.txt};

    			\addplot [lattice data=fitorange]
               	table [x index = 0, y index = 1, x error index = 2, y error index = 3]
               	{FitData/K1points.txt};

               	\addplot[zeroline] coordinates {(-1,0) (170,0)};

                \legend{LO ChPT,LO+NLO ChPT,,,{Data and LO+NLO fit}}
            \end{axis}

        \end{tikzpicture}
    \end{subfigure}
    \\
    \begin{subfigure}[t]{\columnwidth}
    \centering
        \begin{tikzpicture}
            \begin{axis}[
                fit plot,
                xmin=0, xmax=1900, xtick distance=500, restrict x to domain=-1:2600,
                ymin=-3.5, ymax=2, ytick distance=1, restrict y to domain=-4:4,
                xlabel={$(\Mpi/\Fpi)^6$},
                ylabel={$\displaystyle\frac{\KB}{10^3}$}, ylabel shift=0.ex,
                legend pos = south west,
                y tick label style={
                /pgf/number format/.cd,
                fixed,
                precision=0,
                /tikz/.cd
                },
                x tick label style={
                /pgf/number format/.cd,
                fixed,
                precision=0,
                /tikz/.cd
                },
                ]

                \addplot[NLOline, mark=none, filled legend]
                    table[x index=0, y index=1, col sep=tab]
                    {FitData/KBdata.txt};
                \addplot[NLOupperline, mark=none, name path global=NLOU,draw=none]
                    table[x index=0, y index=2, col sep=tab]
                    {FitData/KBdata.txt};
                \addplot[NLOlowerline, mark=none, name path global=NLOL,draw=none]
                    table[x index=0, y index=3, col sep=tab]
                    {FitData/KBdata.txt};

                \addplot [fitgray, opacity=0.3] fill between[of=NLOU and NLOL];

    			\addplot [lattice data=fitorange, legend with mark=fitorange]
               	table [x index = 0, y index = 1, x error index = 2, y error index = 3]
               	{FitData/KBpoints.txt};

               	\addplot[zeroline] coordinates {(0,0) (2100,0)};

                \legend{NLO ChPT,,,,Data}
            \end{axis}
        \end{tikzpicture}
    \end{subfigure}
    \caption[]{
        LO (dashed black line) and NLO (grey line and band) ChPT predictions for $\Kiso$ (top), $\Kisoone$ (middle) and $\KB$ (bottom) as functions of $\Mpi/\Fpi$, using couplings from \cref{eq:LECref}.
        These are compared to lattice results from \rcite{Blanton:2021llb} (orange points) and the best fit to the lattice data (dotted orange line and orange band, not given for $\KB$).
        For reference, the physical point is at \mbox{$(\Mphys/\Fphys)^4\approx5.25$}, \mbox{$(\Mphys/\Fphys)^6\approx12.0$}.}
    \label{fig:fit}
\end{figure}

\Cref{fig:fit} shows the comparison between our results and the lattice data \cite{Blanton:2021llb} at leading ($\Kiso$) and subleading ($\Kisoone$) order in the threshold expansion.
Unlike the LO results, the NLO results are in decent agreement with the data, especially for $\Kiso$.

Data at quadratic order in the threshold expansion is limited to $\KB$, also shown in \cref{fig:fit}.
It agrees poorly with our results; however, this is a very suppressed coefficient, and since it enters first at NLO, it is possible that there are large NNLO corrections.
Overall, we do not have a satisfactory understanding of why NLO corrections are so large, except that NLO introduces qualitatively new contributions such as $\ell^\rr_i$-dependence and the bull's head.
Calculations at NNLO are expected to be very difficult.

\section{Summary and upcoming results}
Our results advance the state of the art in more-than-4-point ChPT scattering, mostly resolve the discrepancy between ChPT and the lattice for $\Kdf$ at maximum isospin, and pave the way for similar studies in other systems.
We also find that the threshold expansion converges well and that the cutoff dependence is small, adding confidence in the validity of the approach.

Work is ongoing to extend our results to general isospin, where the main difficulty is the greater number of channels and the now nontrivial particle exchange (i.e., spectator choice) symmetry, which replaces \cref{eq:threxp} by more complicated threshold expansions.
It is challenging to extract $\Kdf$ from lattice data for non-maximal isospin, and no results are currently available, but we anticipate that they will appear eventually.

Preliminary studies are also ongoing regarding the introduction of kaons and other heavier particles, for which there are some recent results \cite{Draper:2023boj} featuring tension with LO ChPT similar to that we just resolved.
This would require an enhancement of the existing scattering amplitude, which currently supports $3$-flavor ChPT \cite{Bijnens:2022zsq} but not multiple mass scales.








\end{document}